\documentclass{elsart}

\usepackage{graphics}
\usepackage{graphicx}

\usepackage{amssymb}
\usepackage{bm}
\usepackage{color}

\begin{document}

\begin{frontmatter}

\title{Near-threshold measurement of the $\bm{^{4}}$He$\bm{(\gamma,n)}$ reaction}
\vspace*{-10mm}

\author[lund]{B. Nilsson}
\author[lund]{J. -O. Adler}
\author[lund]{B. -E. Andersson}
\author[glasgow]{J. R. M. Annand}
\author[glasgow]{I. Akkurt}
\author[lund]{M. J. Boland}
\author[glasgow]{G. I. Crawford}
\author[lund]{K. G. Fissum\corauthref{cor1}}
\author[lund]{K. Hansen}
\author[glasgow]{P. D. Harty}
\author[glasgow]{D. G. Ireland}
\author[lund]{L. Isaksson}
\author[lund]{M. Karlsson}
\author[lund]{M. Lundin}
\author[glasgow]{J. C. McGeorge}
\author[glasgow]{G. J. Miller}
\author[lund]{H. Ruijter}
\author[lund]{A. Sandell}
\author[lund]{B. Schr\"{o}der}
\author[lund]{D. A. Sims}
\author[glasgow]{D. Watts}

\corauth[cor1]{Corresponding author.  Telephone:  +46 46 222 8618; 
Fax:  +46 46 222 4709; Email address:  \texttt{kevin.fissum@nuclear.lu.se}}

\address[lund]{Department of Physics, University of Lund, SE-221 00 Lund, Sweden}
\address[glasgow]{Department of Physics and Astronomy, University of Glasgow, 
G12 8QQ Glasgow, UK}

\begin{abstract}
\label{section:abstract}
A near-threshold $^{4}$He$(\gamma,n)$ cross-section measurement has been 
performed at MAX-lab.  Tagged photons from 23 $<$ $E_{\gamma}$ $<$ 42 MeV were 
directed toward a liquid $^{4}$He target, and neutrons were detected by 
time-of-flight in two liquid-scintillator arrays.  Seven-point angular 
distributions were measured for eight photon energies.  The results are 
compared to experimental data measured at comparable energies and 
Recoil-Corrected Continuum Shell Model, Resonating Group Method, and recent 
Hyperspherical-Harmonic Expansion calculations.  The angle-integrated 
cross-section data are peaked at a photon energy of about 28 MeV, in 
disagreement with the value recommended by Calarco, Berman, and Donnelly in 
1983.
\end{abstract}

\end{frontmatter}

Over the past several decades, many experiments have been performed in an 
attempt to understand the near-threshold photodisintegration of $^{4}$He.
In 1983, a review article by Calarco, Berman, and Donnelly (CBD)
\cite{calarco83} assessed all available experimental data and made a 
recommendation as to the value of the $^{4}$He$(\gamma,n)$ cross section up
to a photon energy of 50 MeV.  Subsequently, the bulk of the experimental 
effort has been directed towards measuring either the ratio of the 
photoproton-to-photoneutron cross sections or simply the photoproton channel.  
In contrast, only two near-threshold measurements of the photoneutron channel 
have been published \cite{komar93,shima01}.  In this Letter, we report new 
results obtained for the $^{4}$He$(\gamma,n)$ reaction near threshold, and 
compare them with the CBD evaluation as well as the post-CBD data.  We also 
demonstrate consistency with previously published higher-energy tagged-photon 
data \cite{sims98}.  Finally, we compare our data to Recoil-Corrected Continuum
Shell Model (RCCSM) calculations \cite{halderson81,halderson04}, a Resonating 
Group Method (RGM) calculation \cite{wachter88}, and a recent 
Hyperspherical-Harmonic (HH) Expansion calculation \cite{quaglioni04}.  A 
detailed description of the project summarized in this article is given 
in \cite{nilsson03} and will be published in a full article \cite{nilsson05}.

The experiment was performed at the MAX-lab tagged-photon facility 
\cite{adler97}.  A 93 MeV, $\sim$30 nA, pulse-stretched electron beam with a 
duty factor of 75\% was used to produce quasi-monoenergetic photons via 
the bremsstrahlung-tagging technique \cite{adler90}.  Post-bremsstrahlung 
electrons were momentum-analyzed in a magnetic spectrometer equipped with two 
32-counter focal-plane scintillator arrays.  These arrays tagged a 
photon-energy interval from 23 $<$ $E_{\gamma}$ $<$ 42 MeV with a FWHM energy 
resolution of $\sim$300 keV.  The average instantaneous single-counter rate was 
0.5 MHz, and the photon-beam collimation resulted in a tagging 
efficiency \cite{adler97} of $\sim$25\%.

A storage-cell cryostat held the liquid $^{4}$He which constituted the target.  
The cylindrical 75 mm (high) $\times$ 90 mm (diameter) cell of 80 $\mu$m thick 
Kapton was mounted with the cylinder axis perpendicular to the photon-beam 
direction.  The cell was surrounded by a heat shield of three layers of 30 
$\mu$m thick Al foil and multiple layers of the super-insulation NRC-2, all 
maintained at liquid-N$_{2}$ temperature.  The assembly sat in a vacuum 
chamber with 125 $\mu$m thick Kapton entrance and exit windows.  An identical 
empty target cell on the movable target ladder enabled measurement of room and 
non-$^{4}$He background, which turned out to be negligible.  Further, a 1 mm 
thick steel sheet, also mounted on the target ladder, was used to produce 
relativistic $e^{+}e^{-}$ pairs for time-of-flight (TOF) calibration of the 
neutron detectors (see below).  Density fluctuations in the liquid $^{4}$He 
were negligible \cite{tate83}, as was the attenuation of the photon flux due to 
atomic processes within the target materials and the liquid 
$^{4}$He \cite{storm70}.

\begin{figure}[H]
\resizebox{1.00\textwidth}{!}{\includegraphics{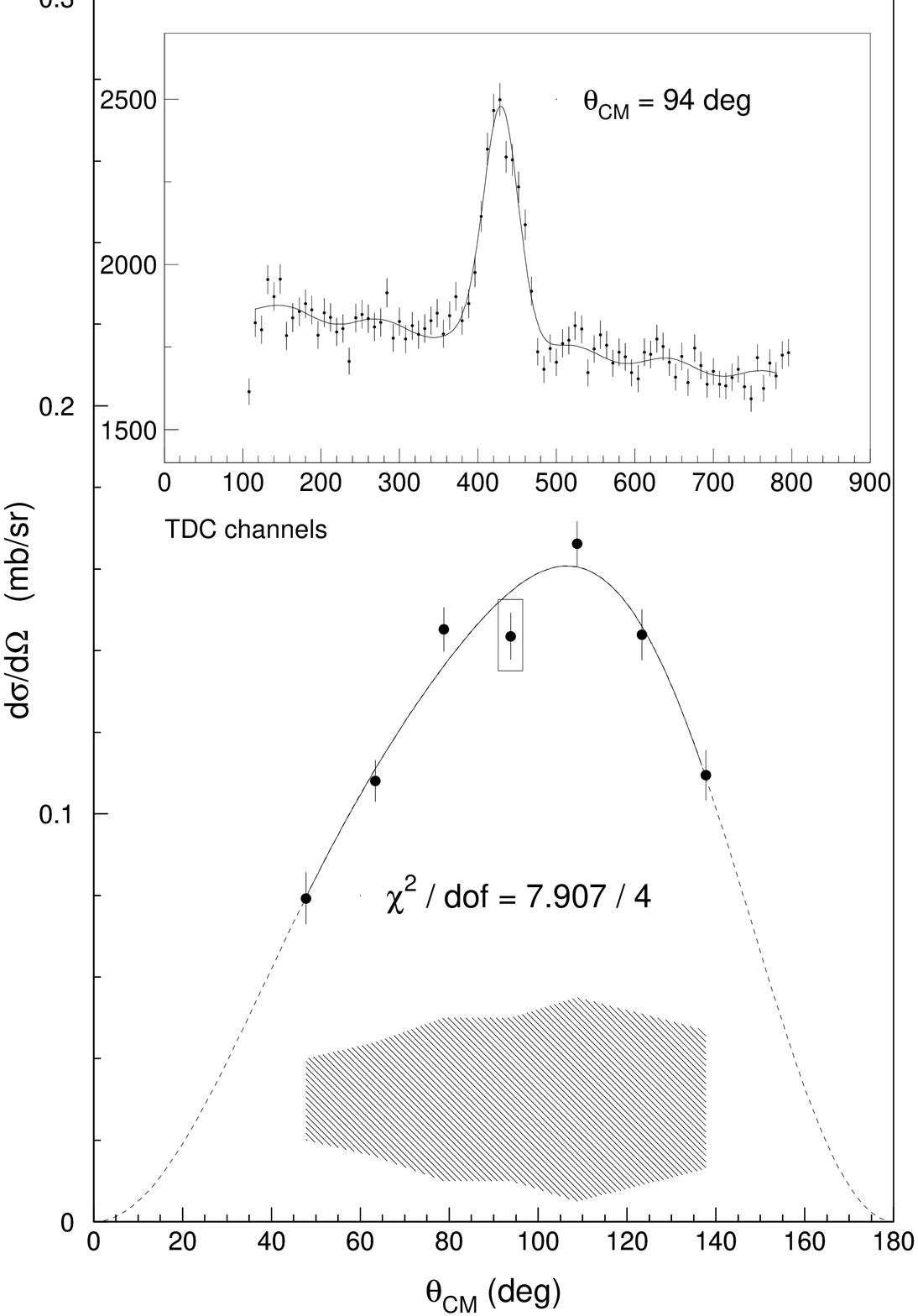}}
\caption{\label{figure:figure_one}
An angular distribution measured at $E_{\gamma}$ $=$ 28.8 MeV.  Error bars are 
the statistical uncertainties, while the systematic uncertainties are 
represented by the band at the base of the panel.  Fitted function (Eq. 
\ref{equation:dsdw_cm_abc}) -- solid line; fitted function extrapolated to 
zero at $\theta_{\rm CM}$ $=$ (0,180) $\deg$ -- dashed line.  The TOF spectrum 
corresponding to the boxed data point at $\theta_{\rm CM}$ $=$ 94 $\deg$ is 
presented in the overset.  The prominent peak corresponds to two-body neutron 
events.  See text for details.}
\end{figure}

Neutrons were detected in two large solid-angle spectrometers \cite{annand97}, 
each consisting of nine 20 cm $\times$ 20 cm $\times$ 10 cm deep rectangular 
cells mounted in a 3 $\times$ 3 lattice and filled with the liquid scintillator
NE213A.  Each of these arrays was mounted on a movable platform (45$^{\circ}$ 
$<$ $\theta_{\rm neutron}$ $<$ 135$^{\circ}$) and encased in Pb, steel, and 
borated-wax shielding.  Plastic scintillators which were 2 cm thick were placed
in front of the liquid scintillators and used to identify incident charged 
particles.  The average flight path to the NE213A arrays was 2.6 m, resulting 
in a 6 msr geometrical solid angle for a single cell and a FWHM TOF 
neutron-energy resolution of $<$2 MeV, which allowed unambiguous identification
of two-body $^{4}$He$(\gamma,n)$ events (see the overset in
Figure \ref{figure:figure_one}).  Thus, the neutron energy also provided a 
cross check on the tagged-photon energy.

Gamma-ray sources were used to calibrate pulse-height output 
\cite{annand97,flynn64,knox72} from the NE213A scintillators which was 
necessary to determine the neutron-detection threshold and thus the 
neutron-detection efficiency.  Pulse-Shape Discrimination (PSD) \cite{annand87}
was employed to distinguish neutrons from photons as the background photon flux
on the TOF spectrometers was $\sim$10$^{5}$ times greater than the neutron 
flux.  All events not seen by the veto detector and identified as neutrons by 
the PSD modules generated a trigger for the data-acquisition system 
\cite{ruijter95}.  The data set for each neutron detector consisted of 64 TOF 
spectra containing real coincidences with the tagger focal plane and a random 
background (see the overset in Figure \ref{figure:figure_one}).  The ratio of
prompt neutrons to random background (due mainly to photons which survived the 
PSD rejection and neutrons resulting from untagged bremsstrahlung) was a strong
function of photon energy, ranging from 1-to-1 at $E_{\gamma}$ $=$ 40.7 MeV to
1-to-10 at $E_{\gamma}$ $=$ 24.6 MeV.  The 64 TOF spectra were summed in eight 
groups of eight tagger counters resulting in $\sim$2.5 MeV wide photon-energy 
bins, each accumulating $\sim$10$^{12}$ photons over the course of the 
measurement.  The background was fitted by superimposing a periodic ripple 
(related to the electron beam circuit time within the pulse-stretcher 
ring \cite{hoorebeke93}) upon an exponential distribution (due to dead-time 
effects in the detectors and the single-hit TDCs used to instrument the focal 
plane).

The background-subtracted neutron yield was corrected for tagger focal-plane 
dead-time effects \cite{hornidge99}.  A \textsc{geant3}-based Monte-Carlo 
simulation \cite{geant93} was used to determine the neutron-yield attenuation 
between the reaction vertex and the detector cells as well as the contribution 
of time-correlated background neutrons scattering into the detectors.  The 
neutron-detection efficiency was determined using the \textsc{stanton} 
Monte-Carlo code \cite{stanton71}.  Cross checks of the predictions made by 
\textsc{geant3} and \textsc{stanton} were performed via a dedicated measurement
of the neutron-detection efficiency using a $^{252}$Cf fission-fragment 
source \cite{karlsson97}.  A summary of the corrections applied to the 
cross-section data and the corresponding systematic uncertainties is presented
in Table \ref{table:table_one}.

The angular distributions measured at each photon energy were converted from 
the laboratory to the Center-of-Mass (CM) frame and fitted
using
\begin{eqnarray}
\label{equation:dsdw_cm_abc}
\lefteqn{
\frac{d\sigma}{d\Omega_{\rm CM}}(\theta_{\rm CM}) = \nonumber } \hspace*{8mm}\\
& & \alpha \left\{\sin^{2}(\theta_{\rm CM})\left[1 + \beta \cos(\theta_{\rm CM}) + \gamma \cos^{2}(\theta_{\rm CM})\right] + \delta + \epsilon \cos(\theta_{\rm CM})~\right\}
\end{eqnarray}
(see Figure \ref{figure:figure_one}).
This expansion assumes that the photon multipolarities are restricted to $E1$, 
$E2$, and $M1$, and that the nuclear matrix elements of the $E$-multipoles to 
final states with a channel spin of unity are negligible\footnote{
Note that Weller {\it\etal~}\cite{weller82} claim non-zero interfering $E1$ 
$S=1$ strength.}\cite{jones91}.  
Under these assumptions, $\alpha$ arises from the incoherent sum of the $E1$, 
$E2$, and $M1$ multipoles, $\beta$ is due to the interference of the $E1$ and 
$E2$ multipoles, $\gamma$ results from the $E2$ multipole, $\delta$ arises from
the $M1$ multipole, and $\epsilon$ is zero.  Similar to analyses of 
complementary $^{4}$He$(\gamma,p)$ angular 
distributions \cite{jones91,calarco03}, our angular distributions were 
constrained to vanish at $\theta_{\rm CM}$ $=$ (0,180) $\deg$ by forcing the 
$\delta$ and $\epsilon$ coefficients to be zero.

\begin{figure}
\resizebox{1.00\textwidth}{!}{\includegraphics{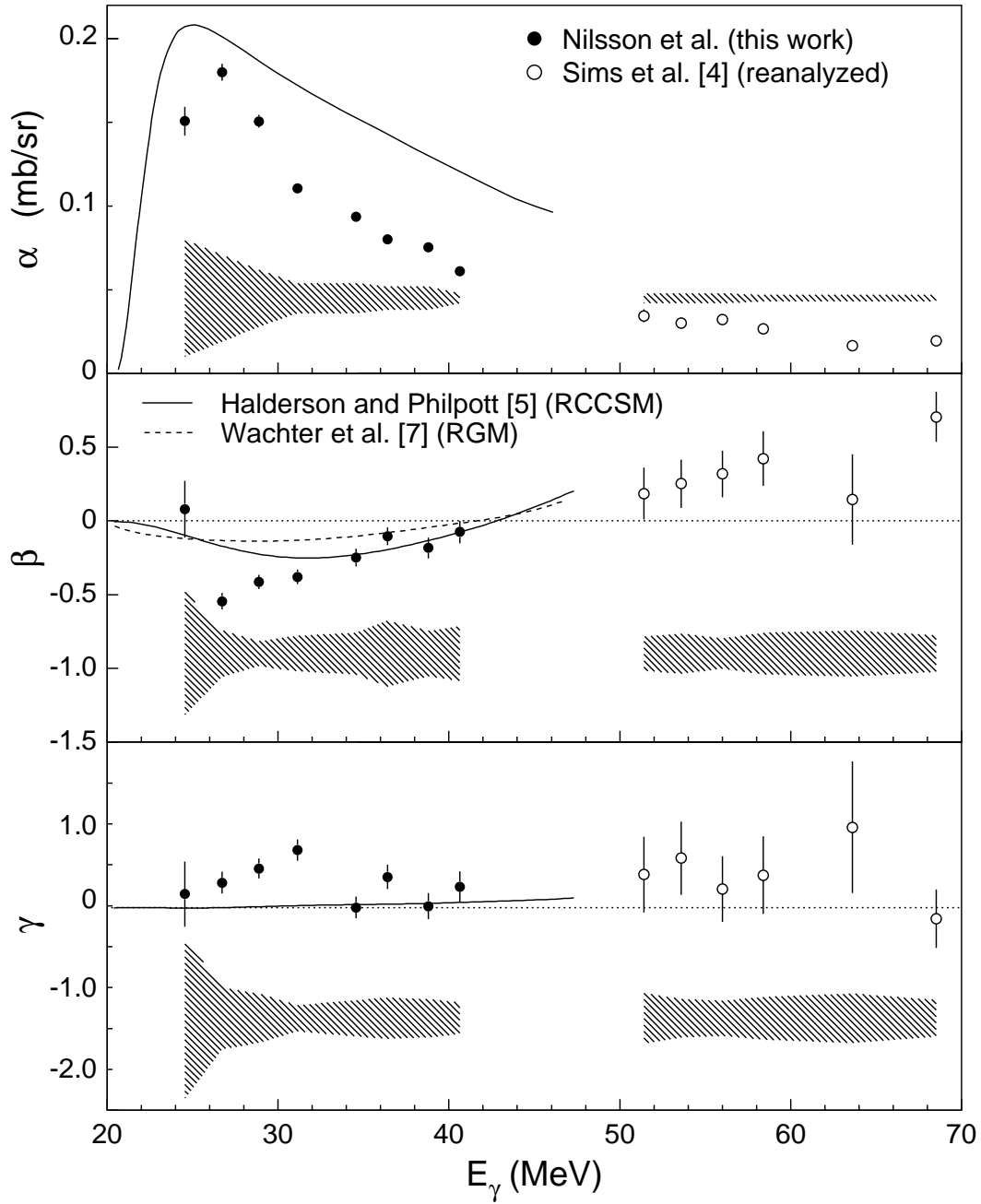}}
\caption{\label{figure:figure_two}
The $\alpha$, $\beta$, and $\gamma$ coefficients:  present data -- filled 
circles; re-analyzed MAX-lab data \cite{sims98,nilsson05} -- open circles; 
RCCSM calculations \cite{halderson81} -- solid lines; RGM calculation 
\cite{wachter88} -- dashed line.  See text for details.}
\end{figure}

Figure \ref{figure:figure_two} presents the $\alpha$, $\beta$, and $\gamma$ 
coefficients (filled circles) together with those extracted from a recent 
reanalysis \cite{nilsson05} of the higher-energy data of Sims {\it\etal~} 
\cite{sims98} (open circles).  We stress that these two data sets from MAX-lab 
are the only tagged-photon data in existence which are differential in angle.  
Error bars are the statistical uncertainties, while the systematic 
uncertainties are represented by the bands at the base of each panel.  Also 
shown are RCCSM \cite{halderson81} and RGM \cite{wachter88} calculations.  The 
recent HH calculation \cite{quaglioni04} does not presently predict angular 
distributions.

The RCCSM calculations were performed within a continuum shell-model framework 
in the ($1p1h$) approximation, where the transition matrix elements of the $M1$
and the spin-independent $M2$ multipole operators vanished.  Corrections were 
applied for target recoil.  In addition to the Coulomb force, the effective 
nucleon-nucleon (NN) interaction included central, spin-orbit, and tensor 
components.  Perturbation theory was employed to compute matrix elements for 
the multipoles and the multipole operators were calculated in the 
long-wavelength limit.  Corrections for spurious CM excitations made these 
calculations essentially equivalent to the multichannel microscopic RGM 
calculations.  Here, a similar semi-realistic NN force was employed, and the 
variational principle was used to determine the scattering wave functions.  The
radiative processes were treated within the Born Approximation, and the 
electromagnetic transition operators were again taken in the long-wavelength 
limit.  Angular momenta up to $L=2$ were allowed in the relative motion of the 
fragments.  Note that the authors of the calculations originally presented 
their results in the form of Legendre coefficients as a function of CM proton
energy.

As can be seen, the data largely reproduce the trends predicted by the 
calculations.  At lower photon energies, the $E1$ multipole is completely 
dominant and the $\alpha$ data have a clear resonant structure peaking at 
about 28 MeV.  The RCCSM calculation tends to overestimate these data, but also
shows resonant structure peaking at about 25 MeV.  The energy dependence of the
$\beta$ data is reasonably consistent with both the RCCSM and the RGM 
predictions, given the relatively large systematic uncertainties for 
$E_{\gamma}$ $<$ 26 MeV.  Similarly, when accuracy and precision are 
considered, there is no significant disagreement between the present $\gamma$ 
data and the RCCSM calculation.  At higher photon energies, $E2$ strength is 
expected to become more important.  Unfortunately, the calculations do not 
cover the range of the higher-energy data.  However, these data do appear to be
consistent with the energy-dependent trends of both the lower-energy data and 
the calculations.

\begin{figure}
\resizebox{1.00\textwidth}{!}{\includegraphics{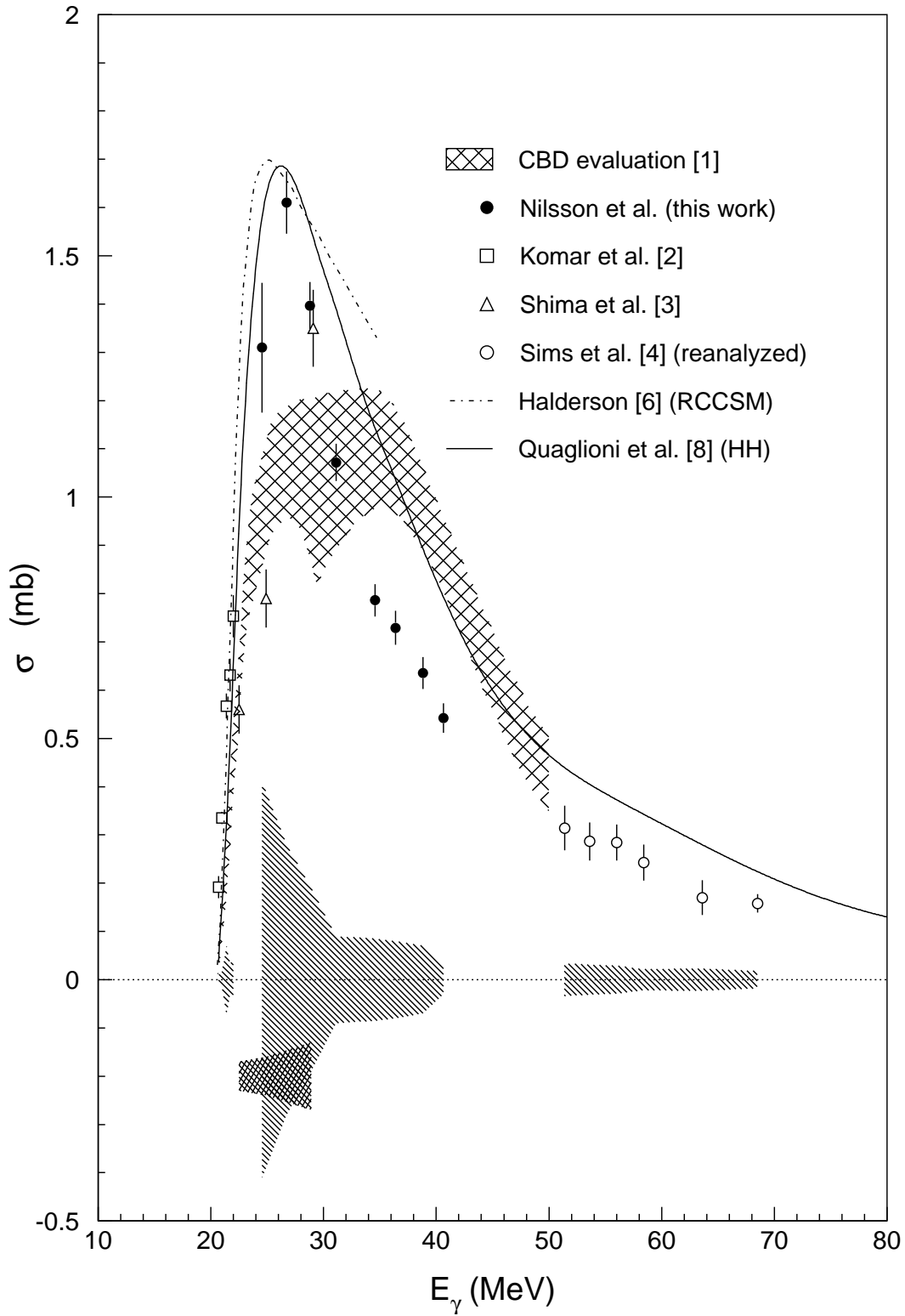}}
\caption{\label{figure:figure_three}
The angle-integrated $^{4}$He$(\gamma,n)$ cross section:  present data -- 
filled circles; re-analyzed MAX-lab data \cite{sims98,nilsson05} -- open 
circles; CBD evaluation \cite{calarco83} -- hatched band; recent RCCSM 
calculation \cite{halderson04} -- dashed-dotted line; and HH calculation 
\cite{quaglioni04} -- solid line.  See text for details.}
\end{figure}

Figure \ref{figure:figure_three} presents the angle-integrated cross-section
data (filled circles) together with those extracted from a recent reanalysis 
\cite{nilsson05} of the higher-energy data of Sims {\it\etal} \cite{sims98}
(open circles).  On average, these angle-integrated data are approximately 7\% 
larger than those which result from simply scaling our $\theta_{\rm CM}$ $=$ 90
$\deg$ results by $8\pi/3$.  Also shown is the CBD evaluation \cite{calarco83},
data from a $^{3}$He$(n,\gamma)$ measurement \cite{komar93}, data from a 
$^{4}$He$(\gamma,^{3}$He$)$ active-target measurement \cite{shima01}, the 
recent RCCSM calculation \cite{halderson04}, and the recent HH 
calculation \cite{quaglioni04}.  Error bars show the statistical uncertainties,
while the systematic uncertainties are represented by the bands at the base of 
the figure.  

The recent RCCSM calculation expanded the model space of Ref. 
\cite{halderson81} to include more reaction channels and all $p$-shell nuclei.
The HH calculation used a correlated hyperspherical expansion of basis states, 
with final-state interactions accounted for using the Lorentz Integral 
Transform Method (which circumvents the calculation of continuum states).  For 
clarity, the small uncertainty in the HH calculation is not shown here.  Note 
that both calculations employ the semirealistic MTI-III 
potential \cite{malfliet69}.

The present $^{4}$He$(\gamma,n)$ excitation function has a clear resonant
structure peaking at about 28 MeV.  Although data are lacking between 42 and 
50 MeV, there is no apparent discontinuity with respect to the re-analyzed 
MAX-lab data of \cite{sims98}.  Furthermore, the present data extrapolate 
smoothly to the lower-energy data of \cite{komar93}.  Conversely, the data of 
\cite{shima01} exhibit a slow rise which is at odds with all other data, 
the calculations, and the CBD evaluation.  Both the RCCSM and HH calculations
are in good agreement with the present data and those of \cite{komar93} up
to the resonant peak at $E_{\gamma}$ $\sim$ 28 MeV.  At higher energies, both
calculations tend to overpredict the cross section, although the HH calculation
follows the general shape of the excitation function up to 70 MeV reasonably
well.  Development of the HH formalism continues \cite{orlandini05}, and we
anticipate new predictions in the near future which use fully realistic NN 
potential models and which may also include 3N-force effects.

In summary, $\frac{d\sigma}{d\Omega}(\theta)$ for the $^{4}$He$(\gamma,n)$ 
reaction have been measured with tagged photons and compared to other available 
measurements and calculations.  The energy dependence of the $\alpha$, $\beta$,
and $\gamma$ coefficients extracted from the angular distributions agrees 
reasonably with trends predicted by RCCSM \cite{halderson81} and RGM 
\cite{wachter88} calculations.  The marked resonant behaviour of the present
angle-integrated cross section, peaking at about 28 MeV, is in good agreement
with recent RCCSM \cite{halderson04} and HH \cite{quaglioni04} calculations
as well as capture data \cite{komar93} which extend close to the $(\gamma,n)$ 
threshold.  This behaviour disagrees with an evaluation of $(\gamma,n)$
data \cite{calarco83} made in 1983, and recent active-target data
\cite{shima01}.

\ack{The authors acknowledge the outstanding support of the MAX-lab staff which
made this experiment successful.  We also wish to thank Sofia Quaglioni,
Winfried Leidemann, and Giuseppina Orlandini (University of Trento, Italy),
John Calarco (University of New Hampshire, USA), Gerald Feldman (The George 
Washington University, USA), Dean Halderson (Western Michigan University, USA),
Andreas Reiter (University of Glasgow, Scotland), and Brad Sawatzky (University
of Virginia, USA) for valuable discussions. BN wishes to thank Margareta 
S\"{o}derholm and Ralph Hagberg for their unwavering support.  The Lund group 
acknowledges the financial support of the Swedish Research Council, the Knut 
and Alice Wallenberg Foundation, the Crafoord Foundation, the Swedish 
Institute, the Wenner-Gren Foundation, and the Royal Swedish Academy of 
Sciences.  The Glasgow group acknowledges the financial support of the UK 
Engineering and Physical Sciences Research Council.}

\begin{small}
\begin{table}[H]
\caption{\label{table:table_one}
A summary of the correction factors applied to the cross-section data and the 
corresponding systematic uncertainties.  In the case of the kinematic-dependent
corrections, average values for the correction and the uncertainty are stated.}
\begin{center}
\begin{tabular}{rcc} \hline \hline
kinematic-dependent quantity & $<$value$>$ & $<$uncertainty$>$ \\
\hline
neutron-detection efficiency &        0.20 &               8\% \\
        neutron-inscattering &        1.25 &               9\% \\
   neutron-yield attenuation &        0.85 &               6\% \\
tagger focal-plane  livetime &        0.95 &               2\% \\
  neutron-detector  livetime &        0.50 &               1\% \\
\hline
              scale quantity &       value &       uncertainty \\
\hline
          tagging efficiency &        0.25 &               3\% \\
  particle misidentification &          -- &               1\% \\
     photon-beam attenuation &          -- &               1\% \\
\hline \hline
\end{tabular}
\end{center}
\end{table}
\end{small}

\end{document}